# "Planck-scale physics" of vacuum in a strong magnetic field

Igor I. Smolyaninov

*Department of Electrical and Computer Engineering, University of Maryland, College Park, MD 20742, USA*

**It is widely believed that Lorentz symmetry of physical vacuum is broken near the Planck scale. Here we show that recently demonstrated "hyperbolic metamaterial" behaviour of vacuum in a strong magnetic field provides us with an interesting analogy of the Planck-scale physics. As demonstrated by Chernodub, strong magnetic field forces vacuum to develop real condensates of electrically charged $\rho$ mesons, which form an anisotropic inhomogeneous superconducting state similar to Abrikosov vortex lattice. As far as electromagnetic field behaviour is concerned, this hyperbolic metamaterial state of vacuum exhibits effective 3D Lorentz symmetry, which is broken at small scale (large momenta) due to spatial dispersion. Thus, an effective Lorentz symmetry-violating "Planck scale" may be introduced. Near the critical magnetic field this effective "Planck scale" is much larger than the metamaterial periodicity defined by the $\rho$ meson lattice. Similar to regular hyperbolic metamaterials, spatial dispersion of vacuum in a strong magnetic field leads to appearance of the "additional wave", which manifests itself as a "heavy" extraordinary photon with an effective mass ~2GeV.**

## 1. Introduction

It is well established that a charged spin-one field interacting with large external magnetic field develops a tachyonic mode if the field magnitude becomes larger than the mass of the spin-one field. This instability of the ground state is resolved by the generation of a vortex state [1-6]. This well-known zero-mode instability has been studied in the literature within the electroweak model in a magnetic field, QCD in a chromomagnetic field, and in colour superconductivity in a magnetic field. Following this line of work, very recently Chernodub demonstrated [7,8] that a strong enough magnetic field forces vacuum to develop real condensates of electrically charged $\rho$ mesons, which form an anisotropic inhomogeneous superconducting state similar to Abrikosov vortex lattice [9] in a type-II superconductor. This unusual effect follows from the same basic consideration of motion of a free relativistic spin $s$ particle in an external magnetic field $B$. The energy levels of the particle are [7,8]

$$E_{n,s_z}^2 = m_\rho^2 + p_z^2 + (2n - 2s_z + 1)|eB| \qquad (1)$$

where $n \geq 0$ is integer, and $s_z=-s, ..., s$ is the spin projection on the field axis, so that the ground state energy of the $s=1$ charged $\rho$ mesons is

$$m_\rho^2(B) = m_\rho^2 - |eB| \qquad (2)$$

The latter result indicates that at large magnetic fields

$$B > B_c = \frac{m_\rho^2}{e} \approx 10^{16} T \qquad (3)$$

vacuum must spontaneously generate positively and negatively charged $\rho$ meson condensates. These condensates appear to be superconducting and spatially inhomogeneous [7,8]. They form a periodic Abrikosov-like lattice of superconducting vortices separated by gaps of the order of $L_B = \sqrt{2\pi\hbar/|eB|} \sim 1\text{fm}$ (Fig.1). Since superconductivity of charged $\rho$ mesons is realized along the axis of magnetic field only, and there is no conductivity perpendicular to magnetic field at *T=0*, strong anisotropy of the vacuum dielectric tensor is observed. The diagonal components of the tensor are positive in the x and y directions perpendicular to the magnetic field, and negative in the z direction along the field. As a result, vacuum in a strong magnetic field behaves as a hyperbolic metamaterial medium [10]. Electrodynamics of hyperbolic media is well studied. Hyperbolic metamaterials exhibit effective 3D Lorenz symmetry [11,12], which is broken at small scale (large wave vectors) due to spatial dispersion [13,14]. Our goal is to apply these general results to the hyperbolic state of vacuum in a strong magnetic field, and study emerging effective "Planck-scale physics" of vacuum.

## 2. Results

First, let us briefly summarize the basic features of hyperbolic metamaterial electrodynamics. These materials belong to the class of uniaxial materials having anisotropic dielectric permittivities $\varepsilon_x = \varepsilon_y = \varepsilon_1$ and $\varepsilon_z = \varepsilon_2$ (here we consider a non-magnetic $\mu = 1$ case). Wave equation for the extraordinary field $\varphi = E_z$ in such a uniaxial material can be written as [10-12]

$$\frac{\partial^2 \varphi}{c^2 \partial t^2} = \frac{\partial^2 \varphi}{\varepsilon_1 \partial z^2} + \frac{1}{\varepsilon_2}\left(\frac{\partial^2 \varphi}{\partial x^2} + \frac{\partial^2 \varphi}{\partial y^2}\right) \qquad (4)$$

While in ordinary crystalline anisotropic media both $\varepsilon_1$ and $\varepsilon_2$ are positive, in the hyperbolic metamaterials $\varepsilon_1$ and $\varepsilon_2$ have opposite signs. These metamaterials are typically fabricated as multilayer metal-dielectric composites or metal wire array structures [9]. In the absence of dispersion, in the case of $\varepsilon_1 > 0$ and $\varepsilon_2 < 0$, eq.(4) would

look like the Klein-Gordon equation for a massless field in a flat (3+1) effective Minkowski space-time:

$$\frac{\partial^2 \varphi}{\varepsilon_1 \partial z^2} = \frac{\partial^2 \varphi}{c^2 \partial t^2} + \frac{1}{(-\varepsilon_2)}\left(\frac{\partial^2 \varphi}{\partial x^2} + \frac{\partial^2 \varphi}{\partial y^2}\right) \qquad (5)$$

However, it is the *z*-coordinate which assumes the role of a time-like variable in this equation. Apparent 4D Lorentz symmetry of this equation obviously violates causality. Therefore, all hyperbolic metamaterials exhibit strong temporal dispersion. On the other hand, spatial dispersion in hyperbolic metamaterials is rather weak. Thus, it is possible to work in the frequency domain around a given frequency $\omega_0$, and write macroscopic Maxwell equations as

$$\frac{\omega^2}{c^2}\vec{D}_\omega = \vec{\nabla} \times \vec{\nabla} \times \vec{E}_\omega \text{ and } \vec{D}_\omega = \vec{\varepsilon}_\omega \vec{E}_\omega \qquad (6)$$

Spatial distribution of monochromatic extraordinary electromagnetic wave inside a hyperbolic metamaterial is described by the following wave equation for $\varphi_\omega$:

$$-\frac{\partial^2 \varphi_\omega}{\varepsilon_1 \partial z^2} + \frac{1}{(-\varepsilon_2)}\left(\frac{\partial^2 \varphi_\omega}{\partial x^2} + \frac{\partial^2 \varphi_\omega}{\partial y^2}\right) = \frac{\omega_0^2}{c^2}\varphi_\omega = \frac{m^{*2} c^2}{\hbar^2}\varphi_\omega \qquad (7)$$

This equation coincides with the 3D Klein-Gordon equation describing a massive scalar field $\varphi_\omega$, in which spatial coordinate *z* behaves as a "timelike" variable. As a result, eq.(7) describes world lines of massive particles which propagate in a flat (2+1) dimensional effective Minkowski spacetime. At small wave vectors *k*, dielectric permittivity components $\varepsilon_1 > 0$ and $\varepsilon_2 < 0$ do not depend on *k*. Therefore, on a large scale hyperbolic metamaterials exhibit effective 3D Lorentz symmetry. This symmetry is broken on a small scale (at large wave vectors) due to spatial dispersion, which manifests itself via weak dependence of $\varepsilon_1$ and $\varepsilon_2$ on the wave vector of the form

$$\varepsilon = \varepsilon^{(0)} + \gamma \left( \frac{k^2 c^2}{\omega^2} \right), \tag{8}$$

where $\gamma$ is small [16]. The dispersion coefficient $\gamma$ can be calculated from the known metamaterial geometry [14,15]. Thus, effective Lorentz symmetry violating "Planck scale" may be written as

$$k_{pl} \approx \frac{\omega_0}{c} \left( \frac{\varepsilon^{(0)}}{\gamma} \right)^{1/2} \tag{9}$$

Small magnitude of $\gamma$ defines smallness of the effective "Planck scale".

Now we can apply these general results (described in detail and verified by numerical simulations in ref. [14-15]) to the hyperbolic metamaterial state of vacuum in a strong magnetic field. In ref.[10] effects of spatial dispersion were neglected, and effective $\varepsilon_{xx} = \varepsilon_{yy} = \varepsilon_1$ and $\varepsilon_{zz} = \varepsilon_2$ permittivities of the hyperbolic state of vacuum were calculated (see eqs.(7-11) in ref.[10]) in the limit $B \approx B_c$ as

$$\varepsilon_2 \approx \alpha \varepsilon_s < 0, \quad \varepsilon_1 = \frac{1+\alpha}{1-\alpha} > 0 \tag{10}$$

where the average volume fraction of the superconducting phase $\alpha$ is small (note that $0 \leq \alpha \leq 1$), and $\varepsilon_s = 1 - \frac{\omega_s^2}{\omega^2} < 0$ is the dielectric permittivity of the superconducting phase ($\omega_s = c/\lambda_L$ where $\lambda_L$ is the London penetration depth [10]). The volume fraction of the superconducting phase can be estimated as [10]

$$\alpha \approx 4\pi r_\rho^3 \rho(B)/3m_\rho, \tag{11}$$

where $r_\rho \sim 0.25$fm is the $\rho$ meson radius, $m_\rho$ is the $\rho$ meson mass, and the $\rho$ meson condensate density is

$$\rho(B) = C_\phi \frac{m_q(B)}{G_V} \left(1 - \frac{B_c}{B}\right)^{1/2}, \qquad \text{at } B > B_c \qquad (12)$$

$$\rho(B) = 0 \qquad \text{at } B < B_c$$

where $C_\phi = 0.51$ is a constant, $m_q(B)$ is the quark mass, and $G_V$ is the vector coupling of four-quark interactions [7,8]. Effects of spatial dispersion in a wire array hyperbolic metamaterial have been considered in great detail by Silveirinha [14]. In the limit of small volume fraction $\alpha$ of the wires, $\varepsilon_1 \approx 1$ does not exhibit any spatial dispersion. On the other hand, $\varepsilon_2$ does exhibit considerable spatial dispersion due to plasma excitations of individual wires [14]:

$$\varepsilon_2 = 1 + \frac{1}{\dfrac{1}{\alpha(\varepsilon_s - 1)} - \dfrac{\omega^2/c^2 - k_z^2}{k_p^2}}, \qquad (13)$$

where

$$(k_p L_B)^2 \approx \frac{2\pi}{\ln\left(\dfrac{L_B}{\pi d}\right) + 0.5275}, \qquad (14)$$

and $d$ is the wire diameter. Since $\varepsilon_s$ is negative and large, in the limit of small $\omega$ and small $k_z$ expression (14) tends to the limiting value defined by eq.(10) from ref.[10]. On the other hand, at large $k_z$ and arbitrary $\omega$ we obtain:

$$\varepsilon_2 = 1 + \frac{1}{\dfrac{1}{\alpha(\varepsilon_s - 1)} - \dfrac{\omega^2}{k_p^2 c^2}} - \frac{k_z^2}{k_p^2 \left(\dfrac{1}{\alpha(\varepsilon_s - 1)} - \dfrac{\omega^2}{k_p^2 c^2}\right)^2}, \qquad (15)$$

which assumes the form of eq.(8). In the limit of small $\omega$

$$\varepsilon_2 = \alpha \varepsilon_s - \frac{k_z^2 \alpha^2 \varepsilon_s^2}{k_p^2}, \qquad (16)$$

and the "effective Planck scale" is

$$k_{pl} \approx \frac{k_p}{|\alpha\varepsilon_s|^{1/2}} \qquad (17)$$

Since $\varepsilon_s$ is negative and very large, Lorentz symmetry violating effects of spatial dispersion show up on a scale, which is considerably larger than $L_B$.

Another interesting consequence of spatial dispersion of the hyperbolic vacuum state is appearance of the additional wave or "heavy photon". This effect is very well known in the physics of materials exhibiting spatial dispersion [16]. Since $\varepsilon_1=1$, a plane wave solution of eq.(7) must satisfy

$$k_\perp^2 = \varepsilon_2\left(\frac{\omega^2}{c^2} - k_z^2\right), \qquad (18)$$

where $k_\perp^2$ is the wave vector component perpendicular to the optical ($z$) axis. Since $\varepsilon_2$ depends on $k_z$, in the presence of spatial dispersion eq.(18) becomes a higher order equation, which may have an "additional wave" solution. Substituting eq.(13) into eq.(18) we obtain

$$k_\perp^2 = \left(1 + \frac{1}{\frac{1}{\alpha\varepsilon_s} - \frac{\omega^2}{k_p^2 c^2} + \frac{k_z^2}{k_p^2}}\right)\left(\frac{\omega^2}{c^2} - k_z^2\right) \approx \frac{(k_p^2 - x)x}{\left(\frac{k_p^2}{\alpha\varepsilon_s} - x\right)}, \qquad (19)$$

where variable $x$ is introduced as

$$x = \frac{\omega^2}{c^2} - k_z^2 \qquad (20)$$

Solving eq. (19) as quadratic equation with respect to $x$, we obtain

$$k_z^2 = \frac{\omega^2}{c^2} - \frac{k_\perp^2 + k_p^2}{2} \pm \sqrt{\frac{(k_\perp^2 + k_p^2)^2}{4} - \frac{k_\perp^2 k_p^2}{\alpha \varepsilon_s}} \qquad (21)$$

At a given small $k_\perp$ eq.(21) has a regular massless solution, which exists at any frequency ω:

$$k_z^{(1)2} \approx \frac{\omega^2}{c^2} - \frac{k_\perp^2 k_p^2}{(k_\perp^2 + k_p^2)\alpha \varepsilon_s} \qquad (22)$$

and the "additional wave" solution, which exists only if $\omega > k_p c$:

$$k_z^{(2)2} \approx \frac{\omega^2}{c^2} - k_p^2 - k_\perp^2 \left(1 - \frac{k_p^2}{(k_\perp^2 + k_p^2)\alpha \varepsilon_s}\right) \qquad (23)$$

This means that its effective mass is

$$m^* = \frac{\hbar k_p}{c} \sim 2 \text{ GeV} \qquad (24)$$

Thus, an additional "heavy" extraordinary photon exists in the hyperbolic metamaterial state of vacuum in a strong magnetic field. Both "regular" and "additional wave" solutions are plotted in Fig.2 for some values of ω. The "heavy photon" is TM polarized, which means that its magnetic field is normal to the axis of DC magnetic field (the z-direction). If we assume that *H* field of the additional wave is directed along the x-direction, its electric field has non-zero components along y- and z-directions. Such "heavy photons" can propagate in any direction inside the hyperbolic metamaterial state. Experimental signatures of such "heavy photons" may help in identifying the hyperbolic metamaterial state of vacuum in accelerator experiments involving heavy ion collisions. According to recent estimates [17], magnetic fields of the required strength $B \sim B_c \sim 10^{16}$ T and geometrical configuration may be created in heavy-ion collisions at the Large Hadron Collider. Based on parity considerations, the preferred channel of "heavy photon" creation would be production of pairs of such photons with the threshold energy of $\sim 2m^*c^2 \sim 4$ GeV. Being collective excitations of the hyperbolic metamaterial state, heavy photons must decay into regular particle jets upon living the $\rho$ meson condensate. Therefore, measured excess of particle jets with ~2 GeV energy may serve as an experimental signature of "heavy photons" and creation of the hyperbolic metamaterial vacuum state.

**3. Conclusions**

In conclusion, we have analyzed the effects of spatial dispersion on the properties of vacuum in a strong magnetic field. As far as electromagnetic field behaviour is concerned, this hyperbolic metamaterial state of vacuum exhibits effective 3D Lorentz symmetry, which is broken at small scale (large momenta) due to spatial dispersion. Therefore, an effective Lorentz symmetry-violating "Planck scale" may be introduced. Near the critical magnetic field, this effective "Planck scale" is much larger than the metamaterial periodicity defined by the Abrikosov lattice of $\rho$ meson condensates. Similar to regular hyperbolic metamaterials, spatial dispersion of vacuum in a strong magnetic field leads to appearance of the "additional wave", which manifests itself as a "heavy" extraordinary photon with an effective mass ~2GeV.

Finally, let us note that quantum effects in the theories with charged fermions (like QED and QCD) in the background of a strong magnetic field are known to lead to appearance of an excitation which looks like a heavy photon [18]. However, these previously known perturbative effects arise from one-loop corrections, and give rise to much lower effective masses ~70 MeV. Unlike these previous results, the effect considered here is different and non-perturbative. The much heavier (~2GeV) TM polarized "heavy photon" state would disappear in the absence of magnetic field induced Abrikosov lattice of $\rho$ meson condensates.

Let us also note that while the anisotropy of the electric response of a fermion system in the presence of a strong magnetic field has been well established in the literature [19-29] (since the external magnetic field breaks the spatial rotational symmetry of the problem, it is natural to expect an anisotropic response to an applied probe electric field), the "hyperbolic" case of extreme anisotropy of vacuum has not been considered previously. As demonstrated above, this "hyperbolic state" of vacuum gives rise to interesting and nontrivial non-perturbative effects.

**Figure Captions**

**Figure 1.** Abrikosov lattice of charged ρ meson condensates forming a "wire array" hyperbolic metamaterial structure of vacuum in a strong magnetic field [10]. The pictorial diagram is drawn schematically to provide visual representation of anisotropic character of superconductivity of the $\rho$ meson condensate: it superconducts only in the direction along magnetic field, while conductivity in the perpendicular direction is zero at T=0.

**Figure 2.** "Regular" (1) and "additional wave" (2) solutions of eq.(19) plotted for some values of the extraordinary photon frequency $\omega$.

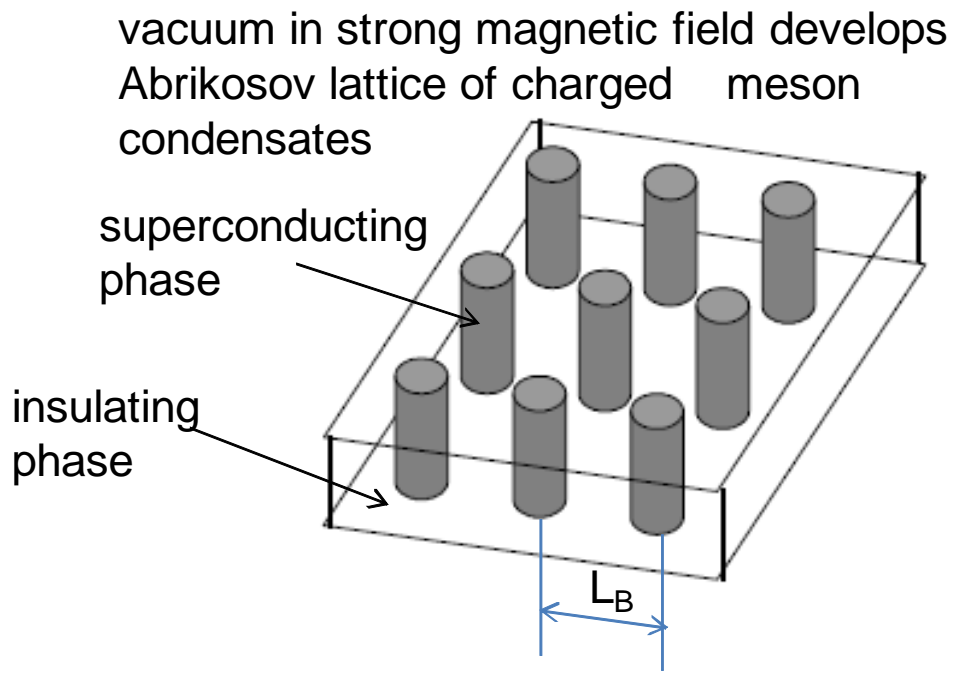

Fig.1

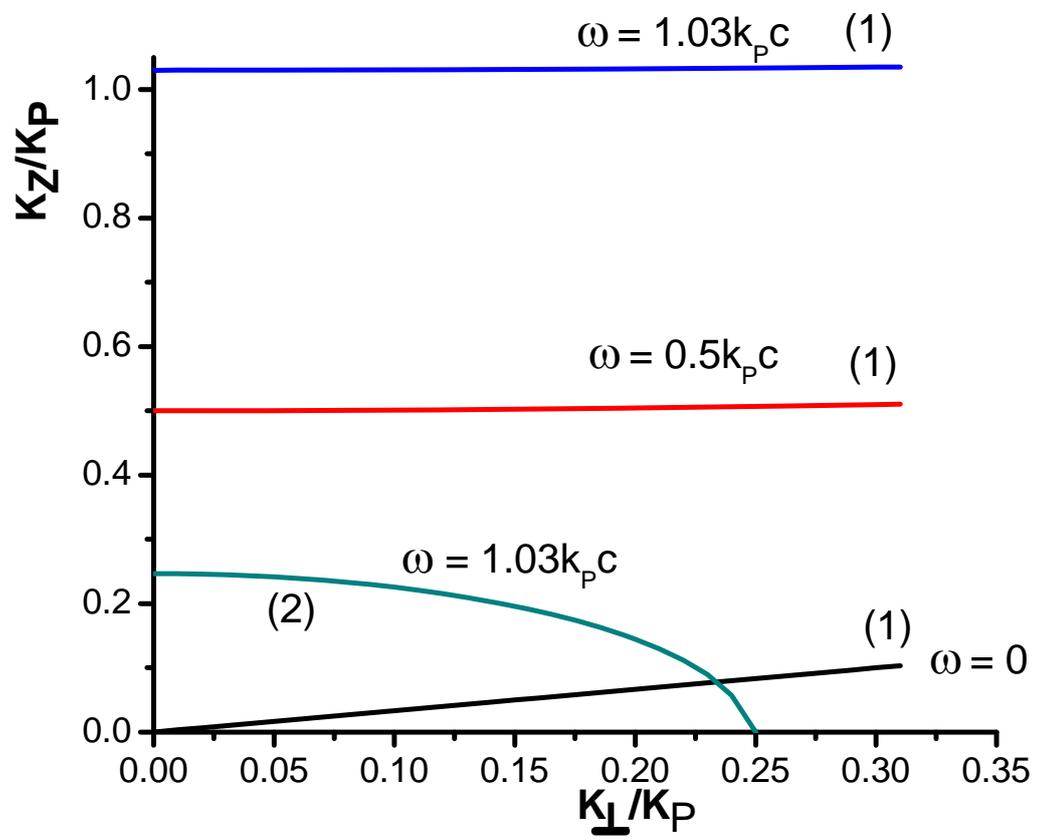

Fig.2